\documentclass[runningheads]{llncs}
\usepackage{graphicx}
\usepackage{proof}
\usepackage{amsmath,amsfonts,stmaryrd,amssymb}
\providecommand\qedhere{\qed}
\usepackage{color}
\usepackage{hyperref}

\newif\iflong

\newcommand\TODO[4][]{\hbox to 0pt{\textcolor[#1]{#2}{$\star$}}\marginpar{\scriptsize\textcolor[#1]{#2}{\textbf{#3} #4}}}

\spnewtheorem{mytheorem}{Theorem}{\bfseries}{\itshape} 
\spnewtheorem{mylemma}[mytheorem]{Lemma}{\bfseries}{\itshape}
\spnewtheorem{myproposition}[mytheorem]{Proposition}{\bfseries}{\itshape}
\spnewtheorem{mysublemma}[mytheorem]{Sublemma}{\bfseries}{\itshape}
\spnewtheorem{mycorollary}[mytheorem]{Corollary}{\bfseries}{\itshape}
\spnewtheorem{myfact}[mytheorem]{Fact}{\bfseries}{\itshape}
\spnewtheorem{mynotation}[mytheorem]{Notation}{\bfseries}{\rmfamily}
\spnewtheorem{myremark}[mytheorem]{Remark}{\bfseries}{\rmfamily}
\spnewtheorem{myexample}[mytheorem]{Example}{\bfseries}{\rmfamily}
\spnewtheorem{myassumption}[mytheorem]{Assumption}{\bfseries}{\rmfamily}
\spnewtheorem{mydefinition}[mytheorem]{Definition}{\bfseries}{\rmfamily}
\spnewtheorem{myrequirements}[mytheorem]{Requirements}{\bfseries}{\rmfamily}
\spnewtheorem{myproblem}[mytheorem]{Problem}{\bfseries}{\rmfamily}
\makeatletter
\newsavebox{\@brx}
\newcommand{\llangle}[1][]{\savebox{\@brx}{\(\m@th{#1\langle}\)}%
  \mathopen{\copy\@brx\kern-0.5\wd\@brx\usebox{\@brx}}}
\newcommand{\rrangle}[1][]{\savebox{\@brx}{\(\m@th{#1\rangle}\)}%
  \mathclose{\copy\@brx\kern-0.5\wd\@brx\usebox{\@brx}}}
\makeatother
\newcommand\up[1]{\overline{#1}}
\newcommand\low[1]{\underline{#1}}
\newcommand\dL{\mathsf{dL}}
\newcommand\keymaerax{\textsc{KeYmaera~X}}
\def\car{\overline{\mathrm{C}}}
\def\cars{\low{\mathrm{C}}}
\def\pos{x}
\def\poss{\low{\pos}}

\newcommand*\diff{\mathop{}\!\mathrm{d}}

\newcommand\mul{\hspace{1pt}{\cdot}\hspace{1pt}}

\def\ti{t}
\def\tis{\low{\ti}}
\def\hpleft{\bigl[}
\def\hpright{\bigr]}
\def\vv{\mathbf{x}}
\def\hp{\alpha}
\def\grammarDef{\ {::=}\ }
\def\amp{\mathbin{\&}}

\def\ode{\D{\vv} = {\ev}}
\def\true{\textsc{true}}
\def\false{\textsc{false}}
\def\formula{\varphi}
\newcommand\Real{\mathbb{R}}

\newcommand\intp[2][]{\bigl\llbracket{#2}\bigr\rrbracket{}_{#1}}
\newcommand\intpp[2][]{\left\llbracket{#2}\right\rrbracket_{#1}}
\newcommand\intpRel[2][]{\mathrel{
	{\relbar}\hspace{-3pt}{\bigl\llbracket}\vcenter{\hbox{$#2$}}
	{\bigr\rrbracket}\hspace{-3pt}{\to}_{#1}}}%\hbox{$\xrightarrow{\hspace{3pt}#1\hfill}$}}}
\def\state{\omega}

\def\solution{\psi}
\def\sol#1{\solution(#1)}

\def\time{t}
\newcommand\Lied[1]{\mathcal{L}_{#1}\,}

\def\tsf{K}

\newcommand\Nat{\mathbb{N}}
\newcommand\Vars{\mathcal{V}}

\newcommand\Terms[1][\Vars]{\mathcal{T}(#1)}
\newcommand\Formulas[1][\Vars]{\mathcal{F}\!\mathit{ml}(#1)}
\newcommand\HP[1][\Vars]{\mathcal{HP}(#1)}
\newcommand\ev{\mathbf{e}}

\newcommand\defeq{\mathrel{\,:=\,}}
\newcommand\AND{\mathrel{\wedge}}
\newcommand\OR{\mathrel{\vee}}
\newcommand\IMP{\Rightarrow}

\newcommand\Deriv[1]{\dot{#1}}%{{#1}'}
\newcommand\D[1]{\dot{#1}}%{{#1}'}

\makeatletter
   \def\@citecolor{blue}%
   \def\@urlcolor{blue}%
   \def\@linkcolor{blue}%

\def\orcidID#1{\smash{\href{http://orcid.org/#1}{\protect\raisebox{-1.25pt}{\protect\includegraphics{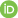}}}}}
\makeatother

\usepackage[firstpage]{draftwatermark}
\SetWatermarkText{\hspace*{10.7cm}\raisebox{16.5cm}
	{\includegraphics{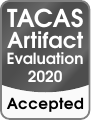}}}
\SetWatermarkAngle{0}

\begin{document}
\title{Relational Differential Dynamic Logic\thanks{Thanks are due to 
Stefan Mitsch, 
Andr\'{e} Platzer, and
Yong Kiam Tan for useful tips on  the KeYmaera X source code; and to 
 Kenji Kamijo, Yoshiyuki Shinya, and Takamasa Suetomi from Mazda Motor Corporation for helpful discussions. The authors are supported by ERATO HASUO Metamathematics for Systems Design Project (No.~JPMJER1603), JST. I.H. is supported by Grant-in-Aid No.~15KT0012, JSPS. J.D. is supported by Grant-in-aid No.~19K20215, JSPS. The work was done during J.K.'s internship at the National Institute of Informatics, Tokyo, Japan.}}
\author{Juraj Kol\v c\'ak\inst{1}\orcidID{0000-0002-9407-9682}\and
J\'er\'emy Dubut\inst{2,3}\orcidID{0000-0002-2640-3065}\and\\
Ichiro Hasuo\inst{2,4}\orcidID{0000-0002-8300-4650}\and
Shin-ya Katsumata\inst{2}\orcidID{0000-0001-7529-5489}\and\\
David~Sprunger\inst{2}\and
Akihisa Yamada\inst{2}\orcidID{0000-0001-8872-2240}}
\authorrunning{J. Kol\v c\'ak et al.}
\institute{LSV, CNRS \& ENS Paris-Saclay, Universit\'e Paris-Saclay, Cachan, France\\
\email{kolcak@lsv.fr}\and
National Institute of Informatics, Tokyo, Japan\\
\email{\{dubut,hasuo,s-katsumata,sprunger,akihisayamada\}@nii.ac.jp}\and
Japanese-French Laboratory for Informatics, CNRS IRL~3527, Tokyo, Japan\and
The Graduate University for Advanced Studies (SOKENDAI), Tokyo, Japan}
\maketitle              % typeset the header of the contribution
\begin{abstract}
In the field of quality assurance of hybrid systems, Platzer's \emph{differential dynamic logic} (dL) is widely recognized as a deductive verification method with solid mathematical foundations and sophisticated tool support. Motivated by case studies provided by our industry partner, we study a \emph{relational extension} of dL, aiming to formally prove statements such as ``an earlier engagement of the emergency brake yields a smaller collision speed.''  A main technical challenge is to combine two dynamics, so that the powerful inference rules of dL (such as the differential invariant rules) can be applied to such relational reasoning, yet in such a way that we relate two different time points. Our contributions are a semantical theory of \emph{time stretching}, and the resulting \emph{synchronization} rule that  expresses time stretching by the syntactic operation of Lie derivative. We implemented this rule as an extension of~\keymaerax, by which we successfully verified relational properties of a few models taken from the automotive domain.
%  relate two states of two dynamics \emph{at different times}. 
% Our main contributions are a theory of \emph{relational differential invariants} (a relational extension of differential invariants in dL) and a concrete technique of \emph{time stretching}. We derive new inference rules for dL from these notions and demonstrate their use in automotive case studies.
%differential equation,
%relational reasoning
\end{abstract}
\newcommand{\myparagraph}[1]{\noindent\textbf{#1}\quad}	
\section{Introduction}
	\label{sec:introduction}
\myparagraph{Hybrid Systems}
\emph{Cyber-physical systems} (CPSs) have been studied as a subject in their own right for over a decade, but the rise of \emph{automated driving} in the last few years has created a panoply of challenges in the quality assurance of these systems. In the foreseeable future, millions of cars will be driving on streets with unprecedented degrees of automation; ensuring the safety and reliability of these automated driving systems is a pressing social and economic challenge.

The \emph{hybridity} of cyber-physical systems, the combination of continuous physical dynamics and discrete digital control, poses unique scientific challenges. To address these challenges, two communities have naturally joined forces: \emph{control theory} whose traditional application domain is continuous dynamics and \emph{formal methods} that have mainly focused on the analysis of software systems. This has been a fruitful cross-pollination: techniques from formal methods such as bisimilarity~\cite{GirardP11} and temporal logic specification~\cite{FainekosP06} have been imported to control theory, and conversely, control theory notions such as Lyapunov functions have been used in formal methods~\cite{TakisakaOUH18}.

\vspace{.3em}
\myparagraph{Deductive Verification of Hybrid Systems}
 In the formal methods community, two major classes of techniques are \emph{model checking} (usually automata-based and automatic) and \emph{deductive verification} (based on logic and can be automated or interactive).
% \footnote{These classes have nonempty intersection: there are some works that aim to integrate (logic- or type system-based) theorem proving and model checking, including~\cite{Kobayashi13}.} 
Model checking techniques rely on exhaustive search in state spaces and therefore cannot be applied \emph{per se} to hybrid systems with infinite state spaces. This has led to the active study of \emph{discrete abstraction} of hybrid dynamics, see e.g.~\cite{GirardP11}; or of \emph{bounded model checking}, see~\cite{dreach}.
% where a principal method is the use of \emph{approximate bisimulations} derived from suitable Lyapunov functions~\cite{GirardP11}.

In contrast,  nothing immediately rules out the use of the deductive approach for hybrid systems. Finitely many variables in logical formulas can represent infinitely many states, and proofs in suitably designed logics are valid even when the semantic domain is uncountable. That said, designing such a logic, proving the soundness of its rules, and showing that logics is actually \emph{useful} in hybrid system verification is a difficult task. 
 % Such a logic should come with concise and intuitive syntax for expressing continuous-time dynamics (i.e.\ differential equations), as well as powerful and versatile reasoning principles for such dynamics. 

 Platzer's \emph{differential dynamic logic} $\dL$~\cite{Platzer18} is a remarkable success in this direction. Its syntax is systematic and intuitive, extending the classic formalism of \emph{dynamic logic}~\cite{HarelTK00} with differential equations as programs. Its proof rules encapsulate several essential proof principles about differential equations, including a \emph{differential invariant} (DI) rule for universal properties and \emph{side deduction} for existential properties. The logic $\dL$ has served as a general platform that accommodates a variety of techniques, including those which come from real algebraic geometry~\cite{PlatzerT18}. Furthermore, $\dL$ comes with sophisticated tool support: the latest tool \keymaerax~\cite{MitschP17} comes with graphical interface for interactive proving and a number of automation heuristics.

 \vspace{.3em}
 \myparagraph{Relational Reasoning on Hybrid Systems}\label{subsec:relational_reasoning_informally}
 \label{subsec:introRelReasoningIndustryPractise}
In this work, we introduce proof-based techniques for \emph{relational reasoning} to
the deductive verification of hybrid systems.
Here, by relational reasoning we mean
analyzing how changes in the system
will affect the overall system behavior.
One of the applications of such reasoning in our mind is to deduce
the safety of a system
by checking the most aggressive settings.
To make such reduction sound, we need to verify that less aggressive versions
result in less dangerous outcomes than the aggressive ones.
As a simple
example, consider the following case distilled from our collaboration with
an industrial partner.

 \begin{myexample}[leading example: collision speed]\label{ex:car}
Consider two cars $\car$ and $\cars$, whose positions and velocities are real numbers denoted by $\up x, \low x$ and $\up v, \low v$, respectively. Their dynamics are governed by the following differential equations:
\begin{equation}\label{eq:leadingExampleODEs}
 \dot{\up x}=\up v,\quad
 \dot{\up v}=1;\qquad\quad
 \dot{\low x}=\low v,\quad
 \dot{\low v}=2.
\end{equation}
Both cars start at the same position at rest ($\up x = \low x = 0 \AND \up v = \low v = 0$), and both drive towards a wall at position $1$. We consider this question: \emph{which car is traveling faster when it hits the wall?}
\begin{displaymath}
\includegraphics[width=.6\textwidth]{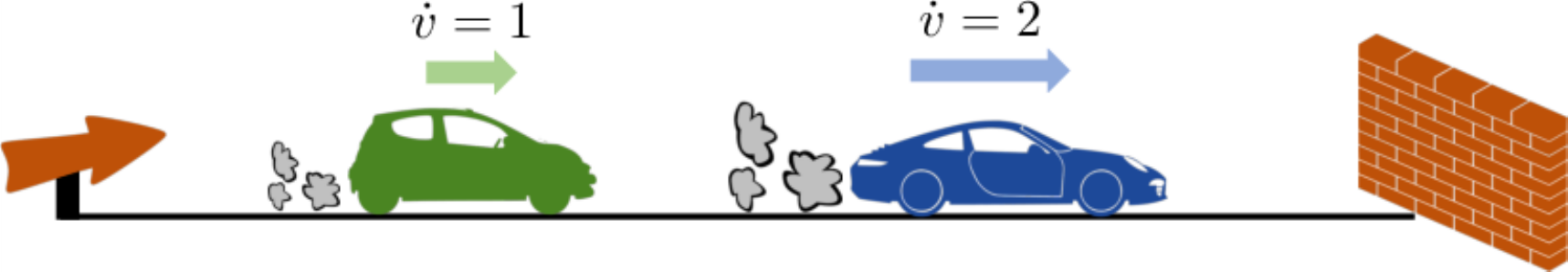}
\end{displaymath}

\begin{figure}[tbp]\centering
\raisebox{-0.5\height}{\includegraphics[width=.35\textwidth]{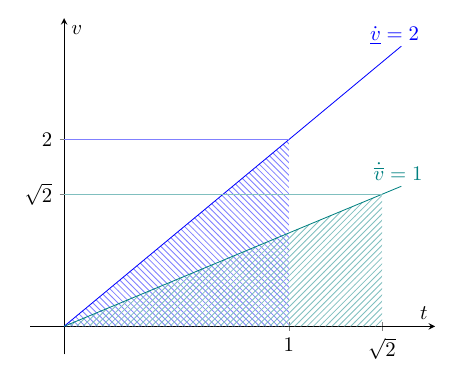}}
 \quad
 \small
 \parbox{.5\textwidth}{ The two hatched areas designate the traveled distances ($\up x=\low x=1$). We can compute the  collision speeds ($\up v=\sqrt{2}$ and $\low v=2$) via the  closed-form solutions of the differential equations~(\ref{eq:leadingExampleODEs}), concluding $\up v \le \low v$ when $\up x=\low x=1$. }
 \caption{An ad-hoc proof for Example~\ref{ex:car}}
 \label{fig:informalProofLeadingExample}
\end{figure}

The second car, $\cars$, has strictly greater acceleration all the time, so we can imagine that $\cars$ hits the wall harder. This hypothesis turns out to be correct, but we are more interested in how this claim could be proven. 

% We would like to be able to prove this claim in a \emph{modular} manner, meaning our arguments should be applicable to other examples of the same nature.

A simple proof would be to solve the differential equation exactly and notice $\cars$ has greater velocity at the end of its run. However, it is known that closed-form solutions are scarce for ODEs---we want a proof method that is more general.

% This kind of proof is feasible due in this scenario to the simplicity of the dynamics. 
% in situations where the dynamics do not admit an explicit solution but the second car still has strictly greater acceleration, we would like to be able to show the second car is moving faster when it hits the wall.

Another possible argument is based on the relationship between the accelerations. Since the second car's acceleration is greater at every point in time, we might be tempted to conclude that the second car's velocity must always be greater than the first car's, based on the monotonicity of integration: $\up a(t) \leq \low a(t) \Rightarrow \up v(t) = \int_0^T\up a(t)\diff t \leq \int_0^T\low a(t)\diff t = \low v(t)$. However, this reasoning has a flaw. $\cars$ reaches the wall at an earlier point in time than $\car$, and therefore $\car$ has more time to accelerate. In the end, we have to compare $\int_0^T \up a(t)\diff t$ and $\int_0^{\low{T}}\low a(t)\diff t$ where $\up a(t) \leq \low a(t)$ for all $t \in [0, \low{T}\,]$ but $T > \low{T}$, as depicted in figure~\ref{fig:informalProofLeadingExample}.

Our solution, roughly stated, is to compare the two cars at the same points \emph{in space} by reparametrizing time for one of the two cars. This parametrization is specially chosen to ensure the two cars pass through the same points in space at the same points in time.

% which enables us to use an argument similar to the one above validly. We will describe this solution more technically in Section 5.
 \end{myexample}

Our current work is about a logical infrastructure needed to support this kind of relational reasoning comparing two different dynamics, based on $\dL$. Our semantical theory, as well as the resulting syntactic extension of $\dL$ by what we call the synchronization rule, generalizes the kind of reasoning in Example~\ref{ex:car} using the notion of \emph{time stretching}. 

% We develop a theory of relational invariants and their connection to simulations between continuous-time dynamics, and embody this theory in the form of concise and useful proof rules in $\dL$.

\vspace{.3em}
\myparagraph{Technical Contributions} 
We make the following technical contributions.
\begin{enumerate}
 \item \textbf{Formulation of relational reasoning in $\dL$}. We find that relational properties are expressible in $\dL$, using disjoint variables in a sequential composition. This representation, however, does not allow the use of the rich logical infrastructure of $\dL$ (such as the (DI) rule). 
 \item \textbf{Time stretching, semantically and syntactically}. To alleviate this difficulty, we first develop the theory  of \emph{time stretching}, so that we can compare two dynamics at different timepoints (cf.\ Example~\ref{ex:car}). Accommodating this semantical notion in $\dL$ and \keymaerax\ is not possible \emph{per se}. We introduce an indirect syntactic alternative, which turns out to be better suited in fact to many case studies (where we compare the two dynamics at the same ``position,'' much like in Example~\ref{ex:car}). The resulting \emph{synchronization} rule in $\dL$ has a clean presentation (Theorem~\ref{thm:sync}), owing to the syntactic Lie derivative operator in $\dL$. 
 \item \textbf{Implementation and case studies}. We implemented the new synchronization rule as an extension of \keymaerax. We used it successfully for establishing nontrivial relational properties in case studies taken from the automotive domain.

 % \item \textbf{Relational invariants and the (RDI) rule.}
 % We define \emph{relational invariants} between two dynamics (Definition~\ref{def:simulation}),
 % give a proof rule (RDI) for these invariants (Definition~\ref{def:simulation_proof_rule}),
 % and formulate a soundness result for this rule (Proposition~\ref{prop:bisimulation_witnesses_RDD_formula}).

 % \item \textbf{Time stretching.} We then turn to particular kinds of 
 % relational invariants induced by reparametrization of time for hybrid
 % systems. These reparametrizations can sometimes be synthesized; we
 % identify some conditions when this can occur and call the resulting
 % reparametrizations \emph{time stretch functions}.

 % % \item \textbf{Additional rules including the (DII) rule.} Besides the two principal new rules (RDI) and (TS), we introduce some $\dL$ rules useful in our case studies (Section~\ref{sec:differential_inductive_invariant}). Notable among these is what we call the different \emph{inductive} invariant  (DII) rule; it is much like the differential invariant (DI) rule in $\dL$ (see e.g.~\cite{PlatzerT18}) but it allows to additionally assume the target inequality itself in an evolution domain constraint.

 % \item \textbf{Case studies} In addition to Example~\ref{ex:car}, we present case studies inspired by our industrial collaboration in Section~\ref{sec:case_study}.
\end{enumerate}

 \vspace{.3em}
 \myparagraph{Relational Reasoning in Practice}
We contend relational reasoning  has practical significance based on our collaboration with an industry partner. Relational properties, especially with an aspect of \emph{monotonicity}, abound in real-world examples. In particular, we have often encountered situations where we have a parametrized model $M(p)$ and need to show a property of the form: 
\begin{equation}\label{eq:monotonicityInformally}
  \text{$p_{1}< p_{2}$ implies $M(p_{2})$ is less safe than $M(p_{1})$. }
\end{equation}
These properties occur  especially in the context of \emph{product lines}, where the same model can come in many slight variants. Example~\ref{ex:car} is such a situation.

Relational statements (such as monotonicity) are easy to state and interpret. Intuitions about the \emph{direction of} the change in a behavior of a system resulting from the change of a parameter are more often valid than intuitions about the \emph{amount of} such a change. These kinds of simple statements are often used by engineers to establish the basic credibility of a model. Qualitative, relational properties also tend to be easier to prove than exact, quantitative properties.

Finally, monotonicity can serve as a powerful technique in \emph{test-case reduction}. If a safety property 
%in a real-world example 
is too complex to be deductively verified, one  usually turns to testing. It is often still possible to establish a simple monotonicity property of the form~(\ref{eq:monotonicityInformally}).
% which must obviously must be weaker than the full safety property. The simple property~(\ref{eq:monotonicityInformally}) 
This can
powerfully boost testing efforts: one can focus exclusively on establishing safety for the extreme case $M(p_{\max})$.

\vspace{.3em}
\myparagraph{Related Work}
Since this work is about its relational extension, the works we
mentioned on $\dL$ are naturally relevant. We discuss other
related works here.

Simulink (Mathworks, Inc.) is an industry standard in modeling hybrid systems, but unfortunately Simulink models do not come with rigorously defined semantics. Therefore, while integration with Simulink is highly desirable any quality assurance methods for hybrid systems, formal verification methods require some work to set up the semantics for Simulink models. The recent work~\cite{LiebrenzHG18} tackles this problem, identifying a fragment of Simulink, and devising a translator from Simulink models to $\dL$ programs. Their translation is ingenious, and their tool is capable of proving rather complicated properties when used in combination with $\keymaerax$~\cite{MitschP17}.

Relational extensions of the \emph{Floyd--Hoare logic}---which can be thought of as a discrete-time version of $\dL$---have been energetically pursued especially in the context of \emph{differential privacy}~\cite{Benton04,AguirreBG0S17,AzevedoGHK19}.

In deductive verification of hybrid systems, an approach alternative to $\dL$ uses \emph{nonstandard analysis}~\cite{Robinson66} and regards continuous dynamics as if they were discrete due to the existence of infinitesimal elements~\cite{SuenagaH11,SuenagaSH13}. The logic used in that framework is exactly the same as the classic Floyd--Hoare logic, and the soundness of the logic in the hybrid setting is shown by a model-theoretic result called the \emph{transfer principle}. Its tool support has been pursued as well~\cite{HasuoS12}.

This is not the first time that relational reasoning---in a general sense---has been pursued in $\dL$. Specifically, Loos and Platzer introduce the \emph{refinement} primitive $\beta\le\alpha$, which asserts a refinement relation between two hybrid dynamics, meaning the set of successor states of $\beta$ is included in that of $\alpha$~\cite{LoosP16}. This kind of relation is inspired by the software engineering paradigm of incremental modeling (supported by languages and tools such as Event-B~\cite{Abrial10EventBBook,ButlerAB16}); the result is a rigorous deductive framework for refining an abstract model (with more nondeterminism) into a more concrete one (with less nondeterminism). In contrast, we compare one concrete model (not necessarily with nondeterminism) with another. Thus, our notion of relational reasoning builds more on relational extensions of the Floyd--Hoare logic~\cite{Benton04,AguirreBG0S17,AzevedoGHK19} than on Event-B. Combining these two orthogonal kinds of relational extensions of $\dL$ is important future work.

\vspace{.3em}
 \myparagraph{Organization} \label{subsec:organization_and_notations}
 In Section~\ref{sec:preliminaries}, we recall some basics of differential dynamic logic $\dL$: its syntax, semantics and some proof rules. 
Our main goal, relational reasoning, is formulated in Section~\ref{sec:rddl}, where we identify difficulties in doing so in the original $\dL$. In Section~\ref{sec:sync} we introduce the semantical notion of time stretching, and turn its theory into the new synchronization rule. After introducing our implementation in Section~\ref{sec:impl}, we describe our three case studies in Section~\ref{sec:case_study}. 
% where we introduce the theory of relational invariants and the (RDI) proof rule. A particularly useful instance of relational invariants arises from the reparametrization of time in hybrid systems; we discuss this \emph{time stretching} in Section~\ref{sec:sync}. In Section~\ref{sec:case_study}, we describe some case studies from the automotive industry.

The appendix containing omitted proofs and details,
the source code and the artifact are found at \url{http://group-mmm.org/rddl_tacas_2020/}.

\section{Preliminaries: Syntax and Semantics of the Logic $\dL$}
  \label{sec:preliminaries}

We recall some of the basics of \emph{differential dynamic logic} ($\dL$).
The interested reader is referred to~\cite{Platzer12,Platzer17} for full details.

\newcommand\deriv[1]{{\textstyle\frac{\partial}{\partial#1}}}

\begin{mydefinition}[language]
We fix a set $\Vars$ of \emph{variables}, denoted by $x,y,\dotsc$.
The set of \emph{terms} is defined by the following grammar:
\[ e, f, g,\dots \; ::=\; x \mid n \mid -e \mid e + f \mid e \mul f \mid e / f
\]
where $x \in \Vars$ and $n \in \Nat$.
First-order \emph{formulas} are defined by
\[
P,Q,\ldots \; ::= \; e \leq f \mid \neg P \mid P \land Q \mid \forall x.\, P
\]

A \emph{state} is a function mapping each variable to a real number,
$\state: \Vars \to {\Real}$. We denote the set of all states by
$\Real^\Vars$. Given a state, each term has a valuation in the reals,
and each formula has a valuation in Booleans defined by the usual
induction. We denote these by $\intp[\state]{e} \in \Real$ and
$\intp[\state]{P} \in \{\true,\false\}$, respectively. The
\emph{models} of a first-order formula $P$ are the states satisfying
$P$, $\intp{P} := \{ \state \in \Real^\Vars \mid
\intp[\state]{P} = \true \}$.
\end{mydefinition}

We use classical shorthands, including $e = f \defeq e \le f \AND f \le e$,
$P \OR Q \defeq \neg (\neg P \AND \neg Q)$, 
$\exists x.\,P \defeq \neg (\forall x.\,\neg P)$, and $\top := 0 \le 0$. 
We denote a vector $(e_1, \dots, e_n)$ of terms (or variables) by $\ev$ when the length $n$ is irrelevant or clear from the context.

  We now introduce the syntax of hybrid programs.
\begin{mydefinition}[hybrid programs]
    \label{def:hybrid_program}
The set $\HP$ of \emph{hybrid programs}
over variables $\Vars$ is given by the following grammar:
    \[
      {\hp}_{1},{\hp}_{2},\dots \;\grammarDef\; {?P} \;\mid\;
x:= e \mid
      \dot x_1 = e_1,\dots,\dot x_n = e_n \amp Q
        \;\mid\; {\hp}_1 ; {\hp}_2  \;\mid\; {\hp}_1 \cup {\hp}_2
 \mid \alpha_1^{*}
    \]
\end{mydefinition}

We may also abbreviate $\D x_1 = e_1,\dots,\D x_n = e_n$
by $\D{\vv} = \ev$.
Hybrid programs of the form $\ode\amp Q$ are especially important in this work. We call such a program \emph{differential dynamics}, where $\ode$ is its \emph{differential equation} and the first-order formula $Q$ is its \emph{evolution domain constraint}.
 The intuitive meaning of such a program is that the values of the variables $\vv$ evolve continuously in time according to $\ode$, as long as $Q$ is satisfied at the current value of $\vv$.
If we see differential dynamics as a continuous analog of loops, then $Q$ plays the role of guard and $\ode$ plays the role of body.\footnote{This analogy is not perfect:
a typical while loop can only exit when its guard is false, whereas a
hybrid program can exit the differential dynamics while $Q$ is satisfied.}
We write $\ode$ instead of $\ode \amp \top$.
% DS: Challenge to find a place where this shorthand is used.
% AY: found \amp \top (instead of \amp \true)

\begin{mydefinition}[solutions]\label{def:solutions}
A mapping $\solution : [0,T) \to \Real^\Vars$ with $T \in [0,\infty]$
is called a \emph{solution} of a differential equation 
$\dot x_1 = e_1,\dots,\dot x_n = e_n$ if $\psi$ is differentiable in 
$[0,T)$ and, whenever $t \in [0,T)$, $\Deriv \psi (t) (x_i) = \intp[\sol{t}]{e_i}$
for $i \in \{1, \ldots, n\}$
and $\Deriv \psi(t)(y) = 0$ for any $y \in \Vars\setminus\{x_1,\dots,x_n\}$.
\end{mydefinition}

According to the Picard--Lindel\"of theorem~\cite{Lindelof94},
for each differential equation $\ode$ and each state $\state$, there is
a unique maximal solution $\solution_\state: [0, T_\state) \to
\Real^\Vars$ of the differential equation satisfying
$\solution_\state(0) = \state$.
%Similarly, for each first-order formula $Q$, there is a
%maximal restriction of that solution obeying $Q$. We
%abuse notation by also denoting it by $\solution_\state: [0,
%T_\state^Q) \to \Real^\Vars$, where
%$T_\state^Q := \displaystyle{}\inf_{t \in [0, T_\state]}
%\solution_\state(t) \notin \intp{Q}$.
%\JD{We may have $\solution_\state(T_\state^Q) \in \intp{Q}$. This happens
%whenever $\intp{Q}$ is closed, e.g., defined with equalities and non-strict 
%inequalities (which is the case in our examples).
%Because of that, I think this formulation is not equivalent to original one...}

\begin{mydefinition}[semantics of hybrid programs]\label{def:sem}
The \emph{semantics} of a hybrid program $\alpha$ is a relation
${\intpRel{\hp}} \subseteq {\Real}^{\Vars} \times {\Real}^{\Vars}$
on states, defined by:
\begin{enumerate}%[label=(\arabic*)]
\item
${\intpRel{?P}} = \{(\state, \state) \mid \state \in \intp{P}\}$,
\item
${\intpRel{x:=e}} = \{(\state, \state') \mid \state'(x) = {\intp{e}}_{\state} 
\text{ and } \state'(y) = \state(y) \text{ for all } y \neq x\}$,
\item
${\intpRel{\D{\vv} = \ev \amp Q}} =
\{(\state,\solution_\state(t)) \mid
  \state \in \Real^\Vars,\ t \in [0,T_\state), \solution_\state([0, t]) \subseteq \intp{Q}\}
$,
\item
${\intpRel{{\hp}_1\cup{\hp}_2}} =
 {\intpRel{{\hp}_1}} \cup {\intpRel{{\hp}_2}}$,
\item
${\intpRel{{\hp}_1; {\hp}_2}} = {\intpRel{{\hp}_1} ; \intpRel{{\hp}_2}}$
where  $;$ denotes relation composition, and
\item 
${\intpRel{{\hp}^*}} = (\intpRel{\hp})^*$
where $^*$ denotes the reflexive transitive closure.
\end{enumerate}
%To be precise:
% where $\solution_\state : [0,T_\state) \to \Real^\Vars$ denotes the unique maximal solution of $\ode$
% such that $\solution_\state(0) = \state$,
% whose unique existence is ensured by the Picard--Lindel\"of theorem.

\end{mydefinition}

\begin{mydefinition}[$\dL$ formulas]
  \label{def:dL_formula}
  \emph{Modal formulas} extend first-order formulas and are defined 
  by the following grammar:
  \[
     \formula,{\formula}_1, {\formula}_2,\ldots \grammarDef
      e \le f \mid \neg\formula \mid
      {\formula}_1 \AND {\formula}_2 \mid \forall x.\,\formula \mid
      \hpleft \hp \hpright \formula.% \mid \langle \hp \rangle \formula
  \]
\end{mydefinition}

As usual, we  write $\langle \hp \rangle \formula$
to abbreviate $\neg \hpleft \hp \hpright \neg \formula$. We will also
call modal formulas ``$\dL$ formulas'' since these are the widest
class of formulas in $\dL$.

The Boolean valuation ${\intp{\formula}}_\state$ of a modal formula
$\formula$ in a state $\state$ is defined in the same way as for
first-order formulas, with the addition of ${\intp{\hpleft \hp \hpright
\formula}}_\state=\true$ if and only if ${\intp{\formula}}_{\state'}=\true$ for all $\state'$ such that $\state
\intpRel{\hp} \state'$.

% \formula}}_\state=\true$ if and only if ${\intp{\formula}}_{\state'}=\true$ for all $\state'$ such that $\state
% \intpRel{\hp} \state'$.

%is
%  the set of states $\state$ such that $(\state,\state') \in \intp{\hp}$
%  implies ${\state}' \in \intp{\formula}$,
% and
%   $\intp{\langle \hp \rangle \formula} =
%     \intp{\neg \hpleft \hp \hpright \neg
%   \formula}$.

%  To ease notation, we write simply $\intp{\formula}$ or $\intp{\hp}$
%  instead of $\intp{\formula}$ or $\intp{\hp}$ respectively,
%  the interpretation $I$ being implicit.
  
% \newcommand\deriv[1]{{\textstyle\frac{\partial}{\partial#1}}}

% Terms are closed under differentiation; \IH{Shall we make this a proper definition?}
% the \emph{derivative}
% of a term $t \in \Terms$ with respect to $x \in \Vars$
% is a term $\deriv{x}e \in \Terms$ which is defined inductively as follows:
% \begin{enumerate}
% \item $\deriv{x}x = 1$ and $\deriv{x}y = 0$ for $y \in \Nat \cup \Vars\setminus\{x\}$,
% \item $\deriv{x}(-e) = -\deriv{x}(e)$,
% \item $\deriv{x}(e_1+e_2) = \deriv{x}e_1 + \deriv{x}e_2$,
% \item $\deriv{x}(e_1\cdot e_2) = \deriv{x}{e_1}\cdot e_2 + e_1\cdot\deriv{x}e_2$,
% \item $\deriv{x}(e_1 / e_2) =
%   (\deriv{x}e_1\cdot e_2 - e_1\cdot\deriv{x}e_2) / (e_2\cdot e_2)$
% \end{enumerate}
% It is easy to see that this syntactic derivative operator is
% semantically correct, in the sense that $\intp{\deriv{x}{e}} =
% \deriv{x}{\intp{e}}$ (as long as both sides are defined).

We take the sequent-calculus style proof system for $\dL$, following~\cite{PlatzerT18}.
It has judgments
of the form $\Gamma \vdash \formula$, where $\Gamma$ is a set of modal
formulas and $\formula$ is a single modal formula.
One of the most fundamental axiom is
\begin{equation}
\hpleft \D \vv = \ev \amp Q \hpright \phi \Longleftrightarrow
\forall t \ge 0.
\left(\forall v \in [0,u].\ [x := f(v)]Q \right) \IMP [x := f(u)]\phi
\tag{solve}
\end{equation}
where $f(t)$ is a term with a fresh variable $t$ such that
$\intp{f}$ is a solution of $\D\vv = \ev$ and
$\intp{f(0)} = \textsf{id}$.

Some other rules of $\dL$, such as the differential invariant rule (DI) that is central in many proofs, are introduced later in Definition~\ref{def:DI}.
  
\section{Relational Differential Dynamic Logic}
\label{sec:rddl}

\newcommand{\Arrow}{\Rightarrow}
\newcommand{\lo}{\low\state}
\newcommand{\uo}{\up\state}
Intuitively,
we want a way to describe two dynamics that are executed in parallel, and compare their outputs.
In terms of (nondeterministic) transition systems,
parallel composition is available via tensor products.
\begin{mydefinition}[tensor product]
 Given two transition systems $(S,R)$ and\linebreak
 $(S',R')$, their {\em
    tensor product} $(S\times S',R\otimes R')$ is defined to be the
  transition system whose transition relation is given by
  \begin{displaymath}
    R\otimes R'\defeq\{(s,s'),(t,t')\mid (s,t)\in R,(s',t')\in R'\}.
  \end{displaymath}
\end{mydefinition}

No extension of the $\dL$ syntax is needed to model tensor products:
disjointness of the variables of the two systems suffices. From now on
we split variables into two disjoint sets: $\Vars = \up{\Vars} \uplus
\low{\Vars}$. We denote variables in $\up\Vars$ by
$\up{x},\up{y},\dots$ and those in $\low\Vars$ by
$\low{x},\low{y},\dots$. Terms in $\Terms[\,\up\Vars\,]$, first-order
formulas in $\Formulas[\,\up\Vars\,]$, and programs in $\HP[\,\up\Vars\,]$ are
denoted by $\up{e},\up{f},\dots$, $\up{P},\up{Q},\dots$, and
$\up{\alpha},\up{\beta},\dots$, and similarly for the corresponding
constructs with $\low\Vars$.

\iflong\else
An easy proof of the following fact can be found in the appendix. 
\fi
\begin{myproposition}\label{prop:monoidal}
${\intpRel{\up\alpha}}\otimes{\intpRel{\low\alpha}} = {\intpRel{\up\alpha;\low\alpha}}$
\qed
\end{myproposition}
\iflong
\begin{proof}
  Note that ${\intpRel{\alpha}} \subseteq \Real^X \times \Real^X$ is defined for any set $X$
  of variables that contains at least all variables occurring in $\alpha$.
  Let us write $\intpRel[X]\alpha$ to make this set $X$ explicit.
  It is easy to see that
  $\state \intpRel[X]{\alpha} \state'$ implies
  $\state(y) = \state'(y)$ for any variable $y \in X$ that does not occur in $\alpha$.
  This means that
  $(\intpRel[\up\Vars\uplus\low\Vars]{\up{\alpha}}) =
   ({\intpRel[\up\Vars]{\up\alpha}} \otimes {=_{\low\Vars}})$
  and
  $(\intpRel[\up\Vars\uplus\low\Vars]{\low{\alpha}}) =
   ({=_{\up\Vars}} \otimes {\intpRel[\low\Vars]{\low\alpha}})$.
  It follows that
  \begin{align*}
  {\intpRel{\up\alpha;\low\alpha}} &=
  {\intpRel[\up\Vars\uplus\low\Vars]{\up\alpha}} \circ {\intpRel[\up\Vars\uplus\low\Vars]{\low\alpha}}\\
  &=
  \Bigl({\intpRel[\up\Vars]{\up\alpha}}\otimes {=_{\low\Vars}} \Bigr) \circ
  \Bigl({=_{\up\Vars}} \otimes{\intpRel[\low\Vars]{\low\alpha}}\Bigr)\\
  &=
  \Bigl({\intpRel[\up\Vars]{\up\alpha}} \circ {=_{\up\Vars}}\Bigr) \otimes
  \Bigl({=_{\low\Vars}} \circ {\intpRel[\low\Vars]{\low\alpha}}\Bigr)\\
  &=
  {\intpRel[\up\Vars]{\up\alpha}} \otimes {\intpRel[\low\Vars]{\low\alpha}}
  \tag*{\qedhere}
  \end{align*}
\end{proof}
\fi

% Since hybrid programs operating on different variables are composed in
% parallel, it does not matter in which order these programs are executed.
% This can be formalized in the $\dL$ logic via the following rule.\IH{It's nice to say what SCC stands for.}

% \begin{myproposition}
% The following rule is sound:
% \[
% \infer[\text{\rm(SCC)}\text{\quad where }\FV{\alpha} \cap \FV{\beta} = \emptyset]{
%   [\alpha][\beta]\phi = [\beta][\alpha]\phi
% }{}
% \]
% \end{myproposition}

Scenarios with two parallel differential dynamics
 are the main focus of this work.
% Recall hybrid programs of the form $\D{x} = e \amp Q$ are
% called differential dynamics. 
We formalize an assertion relating two dynamics using the following format. It is a syntactic counterpart of Proposition~\ref{prop:monoidal}.
% Note that we model the two dynamics, which we assume to evolve in parallel, as a sequential composition. This is a way of accommodating parallel dynamics without changing the $\dL$ syntax. Doing so is justified by (1) enforcing disjointness of variables, and (2) Proposition~\ref{prop:monoidal}. 

% When we have two hybrid programs
% operating in parallel (on disjoint variables), we think of this as a
% relational differential dynamics. \IH{I find the last sentence a bit strange: if variables are disjoint, they are ``unrelated''; still the dynamics is called ``relational''. ... Perhaps can we drop this sentence?} We will be particularly interested in
% certain kinds of formulas about these parallel dynamics, which we
% describe next.

\begin{mydefinition}[relational differential dynamics]
\label{def:rdd}
We call hybrid programs of the following form \emph{relational differential dynamics (RDD)}
\begin{equation}\label{eq:RDD}
  \D{\up\vv} = \up\ev \amp \up{Q}\;;\quad
  \D{\low{\vv}} = \low\ev \amp \low{Q}
\end{equation}
\end{mydefinition}

Now that we have ways to express separate systems evolving in parallel,
we turn to the construction of proofs which reason
about their relationships.

\begin{myexample}\label{ex:simulationForLeadingEx}
Using RDD,
the problem in Example~\ref{ex:car} is expressed as
$
\Gamma_C \vdash
\hpleft\, \up{\delta_C}; \low{\delta_C} \,\hpright \phi_C
$
where
$\up{\delta_C} \defeq \left(\,\D{\up x}=\up v, \D{\up v}=1 \right)$,
$\low{\delta_C} \defeq \left(\D{\low x}=\low v, \D{\low v}=2\right)$,
$\Gamma_C \defeq \{\up x = \low x = 0,\;\up v = \low v = 0 \}$ is the precondition,
and
$\phi_C \defeq (\up\pos=\poss=1 \IMP \up{v} \le \low{v})$ is the postcondition.
% The precondition $\Gamma_C$ asserts that both cars are initially stopped at position $0$,
% and the postcondition $\phi_C$
% asserts that, if one observes the cars at position 1, the velocity of the first car is below that of the second.

Let us prove, in \keymaerax, the RDD sequent
$\Gamma_C \vdash\hpleft\, \up{\delta_C}; \low{\delta_C} \,\hpright \phi_C$.
 In \keymaerax, the only applicable rule to this sequent
 turns it into $\Gamma_C \vdash \hpleft\, \up{\delta_C} \,\hpright \hpleft\, \low{\delta_C} \,\hpright \phi_C$.
We then explicitly ``solve'' the second dynamics, yielding the following goal:
\begin{equation}\label{eq:car solve 1}
	\Gamma_C \vdash \hpleft\, \up{\delta_C} \,\hpright
	\forall \low t \ge 0.
	\left(\up x\, =\, \low x + \low v \mul \low t + \low t^2 = 1
	\;\IMP\; \up v\,\le\, \low v + \low t
	\right)
\end{equation}
where $\low x$ and $\low v$ in $\phi_C$ are replaced by their explicit solutions with respect to the
freshly introduced time variable $\low t$.
Again differential invariant rules do not apply to \eqref{eq:car solve 1},
so one must solve the first dynamics, too, yielding
\[
	\Gamma_C \vdash
	\forall \up t \ge 0.\,
	\forall \low t \ge 0.
	\left(
		\up x + \up v \mul \up t + \up t^2/2\,=\,
		\low x + \low v \mul \low t + \low t^2\,=\, 1\;\IMP\;
		\up v + \up t \,\le\, \low v + \low t
	\right)
\]
Since this goal is first order, the quantifier elimination,
a central proof technique in \keymaerax~\cite{Platzer08},
proves the goal. 
\end{myexample}

The above example worked out since it admits explicit solutions expressible in $\dL$.
This is not always the case as the following example demonstrates.

\begin{myexample}\label{ex:F}
We consider two objects moving through fluids
subjected to different kinds of drag. One object moves through a
viscous fluid and is therefore subject to linear drag; its dynamics are
$\low{\delta_F} \defeq (\dot{\low x} = \low v, \dot{\low v} = -\low v)$.

The other object moves through a less viscous fluid and is subject to
turbulent drag; its dynamics are
$\up{\delta_F} \defeq (\dot{\up x} = \up v, \dot{\up v} = -\up v^2)$.
Our goal is to show that the latter has higher
speed when both objects reach a certain point in space ($\up x = \low x
= l$).

The following functions $\low v^*, \low x^*, \up v^*$ and $\up x^*$ are solutions of the dynamics.
\begin{align*}
\low v^*(\low t) &= \low v_0 \cdot e^{-\low t} &
\low x^*(\low t) &= \low x_0 + \low v_0 \mul (1 - e^{-\low t})
\\
\up v^*(\up t) &= \frac{\up v_0}{1+\up v_0 \mul \up t} &
\up x^*(\up t) &= \up x_0 + \log (1 + \up v_0 \mul \up t)
\end{align*}
where $\low v_0$ etc.\@ denote the initial values.
Unfortunately, we cannot express exponentiations and logarithms in \keymaerax,
and thus the ``solve'' rule that we used in Example~\ref{ex:simulationForLeadingEx} cannot be applied here.
\end{myexample}

One obvious solution to this would be to add support for exponentiations and logarithms in \keymaerax,
but this would break the decidability of the underlying first order logic,
which is a major feature of $\dL$~\cite{Platzer08}.
In fact, the same issue occurs even in standard use cases of \keymaerax,
and motivated the introduction of proof rules which do not demand explicit solutions to differential dynamics~\cite{Platzer17,PlatzerT18}
using the \emph{Lie derivative}.
\begin{mydefinition}[formal Lie derivative in $\dL$ from~\cite{Platzer17,PlatzerT18}]\label{def:LieDeriv}
 The formal Lie derivative of a term $f$ along
 dynamics $\delta \equiv (\D\vv = \ev \amp Q)$ of dimension $n$ is a $\dL$ term
 $\Lied{\delta}f \in \Terms$ given by%
 \footnote{It is easy to see that
 the \emph{derivative}
 of a term $t \in \Terms$ with respect to $x \in \Vars$
 can be given as a $\dL$ term $\deriv{x}e \in \Terms$
 such that
 $\intp{\deriv{x}{e}} = \deriv{x}{\intp{e}}$. The definition of $\deriv{x}{e}$ is inductive with respect to the term $e$. 
 }
 \[
 \Lied{\delta}f := \deriv{x_1}f\cdot e_1 + \dots + \deriv{x_n}f\cdot e_n
 \]
\end{mydefinition}

  \begin{mydefinition}[proof rules from \cite{Platzer17,PlatzerT18}]
    \label{def:DI}
    The following rules are sound:
 % proof rules in $\dL$:
    \[
      \infer[\text{DI}]{\Gamma \vdash \hpleft \delta \hpright f \sim 0}
        {\Gamma, Q \vdash f \sim 0 & \Gamma \vdash
          \hpleft \delta \hpright \Lied{\delta}f \simeq 0}
      \qquad
      \infer[\text{Dbx}]{\Gamma \vdash \hpleft \delta \hpright p \sim 0}
        {\Gamma \vdash p \sim 0 & Q \vdash \Lied{\delta} p \simeq g\cdot p}
    \]
    where
$\delta \equiv (\ode \amp Q)$,
$(\sim, \simeq)\in\{\,(=,=), (>,\ge), (\ge,\ge)\,\}$, and $g$ is any term without division.
  \end{mydefinition}
  The differential invariant rule (DI) is the central rule for proving
  safety properties~\cite{Platzer17,PlatzerT18}: it reduces a global
  property of the dynamics to local reasoning by means of Lie
  derivatives. 
The Darboux inequality rule (Dbx) is derived from real algebraic geometry; see e.g.~\cite{PlatzerT18}.

  \begin{myexample}
   Consider an example differential dynamics in one variable,
$\D{x} = 2$.
Suppose we want to show that
$x \ge 0$ holds after following these dynamics for any amount of time,
starting from $x = 1$. One way to
do this is to show that (1) this predicate holds initially
% ($1 \ge 0$)
and (2) the time derivative of $x$ is always nonnegative. These are
precisely the two premises of the (DI) rule:
to show the sequent
$x = 1 \vdash [\D{x} = 2] x \ge 0$ (DI) requires us to
prove (1) $x = 1 \vdash x \ge 0$ and (2) 
$x = 1 \vdash [\D{x} = 2] \Lied{\D x = 2}x \ge 0$,
where $\Lied{\D x = 2}x = 2$.
Note that
we give an initial condition $x = 1$ in the precedent of this sequent.

  \end{myexample}

\section{Synchronizing Dynamics}
  \label{sec:sync}
The intuitive explanation of the RDD construction of Definition~\ref{def:rdd} is a
``serialization'' of two dynamics.
This construction however does not match the (DI) and (Dbx) rules,
as they accept only one dynamics followed by a comparison.
In order to make use of these rules in our relational reasoning,
we introduce another proof method. It ``synchronizes'' two dynamics.

After some theoretical preparations
we define the new rule and prove its soundness.
We will illustrate the usefulness of
this rule in Section~\ref{sec:case_study}, through some case studies that are inspired by our
collaboration with the industry.

\subsection{Time Stretching}

A key theoretical tool towards the soundness of our synchronization
rule is called \emph{time stretching}. Its idea is very similar to the
technique of \emph{time-reparametrization} for ODEs~\cite{chicone06}.

\begin{mydefinition}[time stretch function]
  \label{def:time_stretch_function}
Let $T\in \Real_{\ge 0}$.
A function $\tsf : [0,T]\to {\Real}_{\geq 0}$ is a \emph{time stretch 
function} if
$\tsf(0) = 0$,
$\tsf$ is continuously differentiable and
$\D{\tsf}(\time) > 0$ for each $\ti\in[0,T]$.
\end{mydefinition}

\begin{myremark}
\label{rem:time_stretch_bijective}
The condition
$\D{\tsf}(\time) > 0$ ensures that 
$\tsf$ is strictly increasing and is a bijection from $[0,T]$ to $[0,\tsf(T)]$.
The inverse of $\tsf$ is $\tsf^{-1} : [0,\tsf(T)] \to [0,T]$, 
and it is straightforward to check $K^{-1}$ is another time stretch 
function.
\end{myremark}

The next results tell us how to turn an ODE into another,
given a time stretching function $K$,
so that
a time-stretch $\psi \circ K$ of a solution $\psi$ of one
becomes a solution of the other.

\begin{mylemma}
\label{lem:tsf}
Suppose $f: \Real^\Vars \to \Real^\Vars$ is a vector field and
$\tsf : [0,T] \to [0,\tsf(T)]$ is a time stretch function.
If $\solution : [0,\tsf(T)) \to \Real^{\Vars}$
satisfies $\Deriv\solution(s) = f(\solution(s))$
for all $s \in [0, \tsf(T))$,
then the function
$\rho = \solution \circ \tsf : [0,T) \to \Real^{\Vars}$
% defined by $\rho(t) := \solution(\tsf(t))$,
satisfies
$\Deriv\rho(t) = \Deriv\tsf(t)\cdot f(\rho(t))$
for all  $t \in [0, T)$.
\end{mylemma}
\begin{proof}
% First, remark that the inverse $\tsf^{-1}$ of $\tsf$
% exists (see Remark~\ref{rem:time_stretch_bijective}).
% We have
We have 
$\Deriv\rho(t)
=\Deriv\tsf(t)\cdot\Deriv\solution(\tsf(t))
=\Deriv\tsf(t) \cdot f(\solution(\tsf(t))) 
=\Deriv\tsf(t) \cdot f(\rho(t))$, 
where the first equality is by the definitions and the chain rule, the second equality is by the assumption on $\Deriv\solution$, and the last equality is by the definition of $\rho$. \qed
% \begin{align*}
% \Deriv\rho(t) &=
% \Deriv\tsf(t)\cdot\Deriv\solution(\tsf(t)) &&\text{(definition and chain rule)}\\
% &= \Deriv\tsf(t) \cdot f(\solution(\tsf(t))) &&\text{(assumption on $\Deriv\solution$)}\\
% &= \Deriv\tsf(t) \cdot f(\rho(t)) &&\text{(definition)}
% \tag*{\qedhere}
% \end{align*}
\end{proof}

Since the inverse of a time stretch function is another time stretch
function, we obtain the following corollary of Lemma~\ref{lem:tsf}.

\begin{mycorollary}
\label{cor:tsf}
Let $\tsf : [0,T] \to [0,\tsf(T)]$ be a time stretch function.
Let $\rho : [0,T) \to \Real^{\Vars}$
satisfy $\Deriv\rho(t) = \Deriv\tsf(t)\cdot f(\rho(t))$
whenever $0 \le t < T$.
Then the function
$\psi : [0,\tsf(T)) \to \Real^{\Vars}$,
defined by $\psi(s) := \rho(\tsf^{-1}(s))$,
satisfies
$\Deriv\psi(s) = f(\psi(s))$
whenever $0 \le s < \tsf(T)$.
\qed
\end{mycorollary}

\subsection{Towards a Syntactic Representation}
\label{sec:syntactic}

So far our time-stretch function $K$ has been a semantical object. Here
we introduce a syntactic way of reasoning via time-stretch functions.
Since a desired time-stretch function is not necessarily expressible in
$\dL$, our syntactic reasoning uses an indirect method that exploits a
pair of functions called a synchronizer. We will be eventually led to a
syntactic reasoning rule (Sync) (Thm.~\ref{thm:sync}).

Given a term $g \in \Terms[X]$ and a mapping $\psi : [0,T) \to \Real^X$,
we define $g_\psi : [0,T) \to \Real$ by
\begin{equation}\label{eq:gpsi}
 g_\psi(t) := \intp[\psi(t)]{g}.
\end{equation}
Intuitively, $g_\psi(t)$ is the value of $g$ at time $t$ when we follow
the dynamics whose solution is $\psi$.

\newcommand\upgs{\up{g}_{\up\psi}}
\newcommand\lowgs{\low{g_\psi}}

\begin{mydefinition}[synchronizers]
\label{def:sync}
Let  $(\up\delta,\low\delta)$ be a  pair of dynamics, 
 $(\up\state,\low\state) \in \Real^{\up\Vars} \times \Real^{\low\Vars}$ be a pair of states, and 
$\up\solution: [0,\up{T}) \to \Real^{\up\Vars}$
and
$\low\solution: [0,\low{T}) \to \Real^{\low\Vars}$ be
the unique solutions of $\up\delta$ and $\low\delta$
from $\up\state$ and
$\low\state$, respectively.
We say a pair of $\dL$ terms $(\up g,\low g) \in \Terms[\,\up\Vars\,] \times \Terms[\,\low\Vars\,]$
\emph{synchronizes}
 $(\up\delta,\low\delta)$ from
 $(\up\state,\low\state)$ 
if the following hold.
\begin{itemize}
 \item $\upgs(0) = \lowgs(0)$ 
 \item The derivatives of
$\upgs$ and
$\lowgs$ are both strictly positive.
\end{itemize}
\end{mydefinition}
%
%Given $\up\ecFun : \Real^{\up\Vars} \to \Real$ and
%$\ecFuns : \Real^{\low\Vars} \to \Real$,
%define the following two functions:
%     \begin{align*}
%      &\up\ecFun_{\up\solution}\colon [0,\up{T})\longrightarrow \Real,
%      &
%      &\up\ecFun_{\up\solution}(\up\ti)=\ecFun(\up\solution(\up\ti));
%      \\
%      &\ecFuns_{\low\solution}\colon [0,\low{T})\longrightarrow \Real,
%      &
%      &\ecFuns_{\low\solution}(\tis)=\ecFuns(\sols{\tis}).
%     \end{align*}
%
The following lemma ensures that,
for any  synchronizer, a corresponding time stretch function exists.

\begin{mylemma}\label{lem:K}
In the setting of Definition~\ref{def:sync},
let  $\up t \in [0,\up T)$
and $\low t \in [0,\low T)$
be such that
 $\upgs(\up\ti) = \lowgs(\tis)$.
Then
the function $K$, defined by $\tsf(s) \defeq \lowgs^{-1}(\upgs(s))$,
is a time stretch function from $[0,\up\ti]$ to $[0,\low\ti]$. 
Moreover we have
$\Deriv\tsf(s) = \frac{\Deriv{\upgs}(s)}{\Deriv{\lowgs}(\tsf(s))}$.
\end{mylemma}

\begin{proof}
Since $\lowgs$ is strictly monotonic on $[0,\tis]$,
it has an inverse
$\lowgs^{-1}$ defined from $\lowgs([0,\tis])$ to $[0,\tis]$.
By assumption we have
$\upgs(0) = \lowgs(0)$,
and thus
$\tsf(0) = \lowgs^{-1}(\upgs(0)) = \lowgs^{-1}(\lowgs(0)) = 0$.
Also since
$\upgs(\up t) = \lowgs(\low t)$, we see that
$\lowgs^{-1}$ is defined from $\upgs([0,\up\ti])$ to $[0,\tis]$.
Thus $\tsf = \lowgs^{-1} \circ \upgs$ is defined from $[0,\up\ti]$ to $[0,\tis]$.
%Furthermore, since $\up\ecFun$ and $\ecFuns$ are continuously differentiable and $\fv$ and
%$\fvs$ are continuous, $\tsf$ is $C^1$.
%Finally,
\begin{align*}
\Deriv{\tsf}(s)
&= \Deriv{\upgs}(s) \cdot \Deriv{\bigl(\lowgs^{-1}\bigr)}(\upgs(s))
&&\text{derivative of }K = \lowgs^{-1} \circ \upgs
\\
&= \frac{\Deriv{\upgs}(s)}{\Deriv{\lowgs}(\lowgs^{-1}(\upgs(s)))}
&&\text{derivative of }\lowgs^{-1}
\\
&= \frac{\Deriv{\upgs}(s)}{\Deriv{\lowgs}(\tsf(s))}
\end{align*}
whose value is positive by assumptions on the derivatives of $\upgs$ and $\lowgs$.
\qed
\end{proof}

We remark that time stretch functions we obtain in Lemma~\ref{lem:K} are not necessarily expressible as a $\dL$ term,
as exemplified by the following example.

\begin{myexample}
Consider two dynamics $\up{\delta_F} \defeq (\D{\up x} = \up v, \D{\up v} = - \up v^2)$
and $\low\delta \defeq (\D{\up x} = 1)$.
Their solutions
$\up\psi, \low\psi : \Real_{\ge0} \to \Real^2$
from initial value $x = 0, v = 1$
are
\begin{align*}
\up \psi(s) &= \left(\log (1 + s), (1 + s)^{-1} \right) &
\low \psi(s) &= (s,0)
\end{align*}
Now let $\up g = \up x$ and $\low g = \low x$.
Then
$\upgs(s) = \log(1+s)$,
$\lowgs = \lowgs^{-1} = \textsf{id}$ and thus $K(s) = \lowgs^{-1}(\upgs(s)) = \log (1 + s)$. This is not rational and not expressible in  $\dL$.
\end{myexample}

Using the syntactic Lie derivative (Definition~\ref{def:LieDeriv}),
we state a sound inference rule that
does not need $\tsf$ to be represented explicitly. We note that there is strong support for
Lie derivatives in the tool \keymaerax, as a key syntactic operation behind
 the differential invariant (DI) rule (Definition~\ref{def:DI}).

\newcommand\sync[1][\up g, \low g]{\otimes_{(#1)}}

\begin{mydefinition}\label{def:synchronizedDyn}
%\IH{Let $(\up g, \low g) \in \Terms[\up\Vars] \times \Terms[\low\Vars]$ be a synchronizer for ... from ...}
%AY: To be precise, being synchronizer has to be proved after using this operation
Let
$\up\delta \defeq \left(\D{\up\vv} = \up\ev\amp\up Q\right)$
and
$\low\delta \defeq \left(\D{\low\vv}=\low\ev\amp\low{Q}\right)$
be two dynamics
and
let $(\up g, \low g) \in \Terms[\,\up\Vars\,] \times \Terms[\,\low\Vars\,]$
(which is supposed to be a synchronizer).
We define the \emph{synchronized dynamics} of $(\up\delta,\low\delta)$ with respect to $(\up g, \low g)$ 
as follows:
\[
\up\delta \sync \low\delta 
\ \defeq\,\Biggl(
\D{\up\vv}=\up\ev,\
\D{\low\vv}=\frac{\Lied{\up\delta}{\up g}}{\Lied{\low\delta}{\low g}}\cdot \low\ev
\Biggr)
\amp \Bigl(\up Q \wedge \low Q \wedge \Lied{\up\delta}{\up g} > 0 
	\wedge \Lied{\low\delta}{\low g} > 0\Bigr)
\]
\end{mydefinition}

\begin{mylemma}
\label{lem:sync}
Let $(\up g,\low g)$ be a synchronizer of $(\up\delta,\low\delta)$
from $(\up\state_0,\low\state_0)$.
 The following are equivalent, where the semantical transition relations are from Definition~\ref{def:sem}.
\begin{enumerate}
 \item \label{item:tr} \begin{math}
(\up\state_0,\low\state_0)
\intpRel{\up\delta;\; \low\delta}
(\up\state,\low\state)
% \in \intp{\up g = \low g}
\end{math}
 and 
\begin{math}
 (\up\state,\low\state)
\in \intp{\up g = \low g}
\end{math}
 \item \label{item:trreci}
\begin{math}
(\up\state_0,\low\state_0)
 \intpRel{\up\delta\sync\low\delta}
 (\up\state,\low\state)
\end{math}
\end{enumerate}
\end{mylemma}
\begin{proof}
We first prove (\ref{item:tr} $\Rightarrow$ \ref{item:trreci}).
% First consider the ``only if'' direction.
In the proof of Lemma~\ref{lem:K},
we can observe that
$\Deriv{\upgs}(s) = \intp[\up\psi(s)]{\Lied{\up\delta}{\up g}}$, and analogously,
$\Deriv{\lowgs}(s) = \intp[\low\psi(s)]{\Lied{\low\delta}{\low g}}$.
Hence we obtain
\begin{equation}\label{eq:K'2}
	\Deriv{K}(s) = \frac{
	\intp[\up\solution(s)]{\Lied{\up\delta}{\up g}}
	}{
	\intp[\low\solution(\tsf(s))]{\Lied{\low\delta}{\low g}}
	} =
	\intpp[\rho(s)]{\frac{\Lied{\up\delta}{\up g}}{\Lied{\low\delta}{\low g}}}
\end{equation}
where $\rho:[0,\up{t}) \to \Real^{\up\Vars\uplus\low\Vars}$ is defined by
$\rho(s) := \left(\up\psi(s),\low\psi(K(s))\right)$.

We note that $K : [0,\up{t}] \to [0,K(\up t)]$ is a time-stretch function, and that  $\low\psi$ is a solution of $\D{\low\vv} = \low\ev$,
that is, $\Deriv{\low\psi}(u) = \intp[{\low\psi}(u)]{\low\ev}$
whenever $0 \le u < \low t = K(\up t)$. Combined with  Lemma~\ref{lem:tsf}, we obtain
% Because of the facts that $K : [0,\up{t}] \to [0,K(\up t)]$ is a time-stretch function,
% from Lemma~\ref{lem:tsf}
% and the fact that $\low\psi$ is a solution of $\D{\low\vv} = \low\ev$,
% that is, $\Deriv{\low\psi}(u) = \intp[{\low\psi}(u)]{\low\ev}$
% whenever $0 \le u < \low t = K(\up t)$,
% we know that
\begin{align*}
\Deriv{\bigl(\low\psi\circ K\bigr)}(s) &=
\Deriv\tsf(s)\cdot \intp[\low\psi(K(s))]{\low\ev} =
\Deriv\tsf(s)\cdot \intp[\rho(s)]{\low\ev}&&
\text{ whenever }0 \le s < \up t.
\end{align*}
Hence,
with the fact that $\up\psi$ is a solution of $\D{\up\vv}=\up\ev$,
we obtain
\[
\Deriv{\rho}(s) =
\left(\Deriv{\up\solution}(s),\Deriv{\bigl(\low\solution\circ K\bigr)}(s)\right) =
 \left(
\intp[\rho(s)]{\up\ev},\ \Deriv\tsf(s)\cdot\intp[\rho(s)]{\low\ev}
\right) =
\intp[\rho(s)]{\Bigl(
\up\ev,\ \frac{\Lied{\up\delta}{\up g}}{\Lied{\low\delta}{\low g}}\cdot \low\ev
\Bigr)
}
\]
whenever $0 \le s < \up t$.
Here the last equality is from \eqref{eq:K'2}.
This concludes that $\rho$ is a solution of the dynamics $\up\delta \sync \low\delta$.
It remains to prove that for all $\tau \in [0,\up t]$, 
$\intp[\rho(\tau)]{\up Q \land \low Q \land \Lied{\up\delta}{\up g} > 0 
	\land \Lied{\low\delta}{\low g} > 0}$ is true.
This is an easy consequence of  item~\ref{item:tr}, and the fact that $(\up g,\low g)$ 
is a synchronizer of $(\up\delta,\low\delta)$
from $(\up\state_0,\low\state_0)$.

For the direction (\ref{item:trreci} $\Rightarrow$ \ref{item:tr}), let 
$(\up\xi,\low\xi):[0, T)\rightarrow\mathbb{R}^{\up\Vars}\times\mathbb{R}^{\low\Vars}$
be the unique solution of $\up\delta\sync\low\delta$ from $(\up\state_0,\low\state_0)$.
Then there is $t \in [0,T)$ such that $(\up\xi(t),\low\xi(t)) = (\up\state,\low\state)$.
Let us prove that $(\up\state,\low\state) \in \intp{\up g = \low g}$.
The function $h: s\in[0,T)\mapsto\intp[\up\xi(s)]{\up g}-\intp[\low\xi(s)]{\low g}$
is equal to $0$ at $s=0$ and its derivative is given by:
$$\Deriv h(s) = \intp[\up\xi(s)]{\Lied{\up\delta}{\up g}}-
	\intp[\low\xi(s)]{\Lied{\low\delta}{\low g}}.
		\frac{\intp[\up\xi(s)]{\Lied{\up\delta}{\up g}}}
			{\intp[\low\xi(s)]{\Lied{\low\delta}{\low g}}}=0$$
Consequently, $h$ is the constant function equal to $0$, which implies that 
$(\up\state,\low\state) \in \intp{\up g = \low g}$.
By definition, $\up\xi$ is a solution of $\up\delta$, so 
$\up\state_0 \intpRel{\up\delta} \up\state$. 
Furthermore, by Corollary~\ref{cor:tsf}, $\low\xi\circ K^{-1}$ is a solution of $\low\delta$.
Thus $\low\state_0 \intpRel{\low\delta} \low\state$ and 
\begin{equation*}
(\up\state_0,\low\state_0) \intpRel{\up\delta;\;\low\delta} (\up\state,\low\state)
\tag*{\qedhere}.
\end{equation*}
\end{proof}

% The above lemma states that,
% if the relational dynamics
% are entered and exited while $\intp{\up g = \low g}$ is satisfied,
% then they can be synchronized.\IH{Is it OK to add ``using  $(\up g,\low g)$ as a synchronizer''?}
% Therefore, we need to ensure that initial conditions ensure $\up g = \low g$,
% and that postconditions assume $\up g = \low g$.
% To be more general, we consider postconditions of form $E \IMP \phi$
% such that $E$ implies $\up g = \low g$.
% These observations lead us to the following rule. It allows us to \emph{synchronize} the relational dynamics $\up\delta; \low\delta$
% as a unified dynamics $\up\delta\sync\low\delta$.
The above lemma is a key observation in the current work. It allows us to turn the relational dynamics $\up\delta; \low\delta$---expressed as a sequential composition in $\dL$---into a combined dynamics $\up\delta\sync\low\delta$. Moreover, we can do so in a way that the two dynamics are synchronized in a reparametrized manner, as specified by $(\up g,\low g)$. Such combination of two dynamics is crucial in exploiting the logical infrastructure of $\dL$ and \keymaerax---we emphasize again that the (DI) rule does not support invariant reasoning about the relationship between $\up\delta$ and $\low\delta$, when the relational dynamics is expressed in the original form $\up\delta;\; \low\delta$. 

The following is an incarnation of Lemma~\ref{lem:sync} as a proof rule.  We assume that a postcondition is a conditional form $E\IMP \varphi$; $E$ is called an \emph{exit condition}. By assuming that $E$ implies $\up g = \low g$, we enforce the second condition $ (\up\state,\low\state)
\in \intp{\up g = \low g}
$ in item~\ref{item:tr} of Lemma~\ref{lem:sync}. The first three premises are there to ensure that $(\up g, \low g)$ is a synchronizer.  Under these premises (the first four), the rule allows one to transform its conclusion (about $\up\delta; \low\delta$) into one about the combined dynamics $\up\delta\sync\low\delta$, which is amenable to application of the (DI) rule, for example.
\begin{mytheorem}[synchronization rule]\label{thm:sync}
The following inference rule is sound:
\[
\infer[(\text{Sync})]{
\Gamma \vdash [\,\up\delta; \low\delta\,] (E \IMP \varphi)
}{
\Gamma \vdash [\,\up\delta\,] \Lied{\up\delta}{\up g} > 0\hfil&
\Gamma \vdash \up g = \low g\hfil\cr
\Gamma \vdash [\,\low\delta\,] \Lied{\low\delta}{\low g} > 0\hfil&
E \vdash \up g = \low g\hfil&
\Gamma \vdash [\,\up\delta\sync\low\delta\,](E \IMP \varphi)
}
\]
%where
%$\up\delta \equiv (\D{\up\vv}=\up\ev \amp \up Q)$ and
%$\low\delta \equiv (\D{\low\vv}=\low\ev \amp \low Q)$.
\end{mytheorem}
Recall the definition of $\up\delta\sync\low\delta$ (Definition~\ref{def:synchronizedDyn}), where  time stretching for the second dynamics $\low\delta$ is expressed syntactically by Lie derivatives. 
We call the four premises
$\Gamma \vdash \up g = \low g$,
$E \vdash \up g = \low g$,
$\Gamma \vdash [\,\up\delta\,] \Lied{\up\delta}{\up g} > 0$, and
$\Gamma \vdash [\,\low\delta\,] \Lied{\low\delta}{\low g} > 0$
the \emph{synchronizability conditions}.
These obligations are usually easy to discharge.
The last premise,
which we call the \emph{synchronized formula},
is typically the core remaining obligation.

\begin{myremark}[choice of  $(\up g, \low g)$]
  In applying the (Sync) rule, one still has to find a suitable synchronizer $(\up g, \low g)$. This turns out to be straightforward in many examples.  In all the case studies in Section~\ref{sec:case_study} and in Example~\ref{ex:car}, the exit condition $E$ is of the form $\up x = \low x = C$ where $C$ is a constant. This suggests the use of $\up g = \up x$, $\low g = \low x$. Indeed, all our proofs use this choice of $(\up g, \low g)$.
\end{myremark}

\section{Implementation}
\label{sec:impl}
{\keymaerax} {\cite{keymaeraxurl}} is an interactive theorem prover based on the
sequent calculus formulation of dL. It is implemented in Scala, replacing its
former system KeYmaera {\cite{keymaeraurl}}. It has a web-based GUI environment, and a
support of automated theorem proving using computer algebra systems such as
Mathematica~\cite{Mathematica}.

For the formalization of case studies in Section~\ref{sec:case_study},
we extended \keymaerax\ version 4.7 (available at {\cite{keymaeraxurl}})
with the (Sync) rule. This extension of KeYmaera X,
together with our proofs in case studies, are currently available at
\url{http://group-mmm.org/rddl_tacas_2020/}.

The \keymaerax~implementation is structured in a flexible manner, from which we benefited.
To add  a rule to \keymaerax, one  has to implement  a Scala
program that take the conclusion of the rule and generate the
premises of the rule as subgoals. The fact that any Scala program is allowed here enabled us to implement  complex algorithms, such as inductive translation of formulas.
% To add  rules to \keymaerax, one  implemented  Scala
% programs that take the conclusion of the rule and generate the
% premises of the rule as subgoals.
% Inside the program, we can
% write complex algorithms, such as inductive translation of formulas and
% computing Lie derivatives of expressions. 

% To add a rule to \keymaerax,
% we follow the implementation of existing rules:
% we implemented a Scala function that takes the conclusion sequent of the rule and generates
% premise judgements of the rule as subgoals.
In implementing the (Sync) rule, the functions in~\keymaerax\ called \emph{helpers} helped us,
such as in the Lie derivative computation and the functionality to simplify formulas into equivalent ones.
The bulk of our effort regarded the $\sync$ operator.
There we did a bit more general than we stated in the paper:
not only taking dynamics of form $\ode\amp Q$,
we also allow sequences of dynamics possibly interleaved by guards and nondeterministic choices. This feature was utilized in the case study that will be described in Section~\ref{sec:switch}.

% Our modifications are located in
% \verb|/keymaerax-core/src/main/scala/edu/cmu/cs/ls/keymaerax/btactics/|
%
% The programs corresponding to
% the (Sync) are written as subclasses of
% \verb!RightRule! class, then packed into a file and placed in
% \verb!btactics! directory inside \verb!keymaerax-core! source
% tree. These rules are registered to \keymaerax\ as tactics by adding extra
% codes in \verb!Axiominfo.scala!. At the same time, we implemented test
% codes for these rules; they are placed in btactics directory inside
% \verb!keymaerax-webui! source tree.

\section{Case Studies}
\label{sec:case_study}

We describe three case studies where we proved
relational properties of hybrid dynamics. We did so
formally in our extension of \keymaerax\ described in Section~\ref{sec:impl}.
In all the examples, we apply the (Sync) rule as a main proof step,
in conjunction with the existing rules in $\dL$.
Below, we
describe our example systems and outline the important steps in the formal proofs.

\subsection{Collision Speed with Constant Acceleration}
\label{subsec:const}
In this section we apply the (Sync) rule to the running Example~\ref{ex:car}.
For this example we consider two dynamics
$\up{\delta_C} \defeq \left(\D{\up x} = \up v, \D{\up v} = \up a\right)$ and
$\low{\delta_C} \defeq \left(\D{\low x} = \low v, \D{\low v} = \low a\right)$.
Both dynamics represent a car with constant acceleration.
Our claim is that if acceleration is larger in the first system,
then the first car is necessarily faster than the second car after traveling the same distance $l$;
formally,
\begin{align}
\label{eq:constant_acceleration_formula}
	\Gamma \vdash \hpleft\,\up{\delta_C}; \low{\delta_C} \,\hpright
	(\up x = l \land \low x = l \IMP \low v \leq \up v)
\end{align}
where
\[
\Gamma \defeq \{ 0 = \up x = \low x 
,\ \ 0 < \up v = \up v_0
,\ \ \low v = \low v_0
,\ \ \up v_0 \geq \low v_0
,\ \ 0 \le \low a \le \up a\}
\]

We apply the (Sync) rule, where $\up g \defeq \up x$ and $\low g \defeq \low x$.
The first two synchronizability conditions are
$\Gamma \vdash \low x = \up x$ and
$\low x = l,\ \up x = l \vdash \low x = \up x$, which are trivial.
The last two synchronizability conditions are
\begin{align*}
\Gamma &\vdash \hpleft\, \up{\delta_C}\,\hpright \Lied{\up{\delta_C}}{\up g} = \up v > 0
&
\Gamma &\vdash \hpleft\, \low{\delta_C}\,\hpright \Lied{\low{\delta_C}}{\low g} = \low v > 0
\end{align*}
which are proven using differential invariants (DI). The synchronized formula is
\[
\Gamma \vdash \hpleft\, \up{\delta_C}, 
\D{\low x} = \low v \mul (\up v / \low v), \D{\low v} = \low a \mul (\up v/\low v)
\amp \up v > 0 \land \low v > 0
\hpright (\up x = l \wedge \low x = l \IMP  \low v \leq \up v)
\]

One might try to show the inequality $\up v - \low v \geq 0$ by the
differential invariant (DI) rule, but the Lie derivative of the term $\up v - \low v$
is $\up a - \low a \cdot(\up v/\low v)$, which is not obviously
nonnegative.
Instead,
a trickier expression
$\up a \cdot (\low v^2 - \low v_0^2) - \low a \cdot (\up v^2 - \up v_0^2) = 0$
turns out to be an invariant. Its Lie derivative
is $\up a \cdot (2\low v) \cdot \low a\cdot (\up v/\low v) - \low a \cdot(2
\up v)\cdot \up a$, which is clearly 0, since we also know $\low v >
0$.

We do not have an intuitive explanation for this invariant, but it was
found by a template-based search, like many other invariants in $\dL$.
By positing the existence of a polynomial invariant of a
certain degree, we can find conditions on the coefficients by requiring
its Lie derivative and initial value are zero. Solving these conditions
for a second-degree invariant on the velocities in the system yielded
the invariant above.

After finding our invariant, we additionally have to show the invariant
entails our desired result, $\low v\le \up v$. This can be shown with a
standard monotonicity property of modal logics: from $\phi \vdash \psi$
and $\Gamma \vdash [\alpha] \phi$, we can conclude $\Gamma \vdash
[\alpha] \psi$, where $\phi$ states the expression above is an
invariant and the velocities are always greater than their initial
value, and $\psi$ is our goal: $\low v \le \up v$.

\subsection{Collision Speed with Different Kinds of Friction}

Here we continue Example~\ref{ex:F},
where we consider two dynamics
$\up{\delta_F} \equiv (\dot{\up x} = \up v, \dot{\up v} = -\up v^2)$ and
$\low{\delta_F} \equiv (\dot{\low x} = \low v, \dot{\low v} = -\low v)$.
Our goal is $\Gamma_F \vdash
[\,\up{\delta_F};\low{\delta_F}\,] (\up x = \low x = l \IMP \low v \le \up v)$, with
$\Gamma_F \defeq \{\up x = \low x = 0,\ 0 < \low v \le \up v \le 1\}$.

First, we establish the fact that the objects in this example always
have positive velocity.
We show this by the (Dbx) rule (Definition~\ref{def:DI}),
where $\Lied{\up{\delta_F}}{\up v} = -\up v^2$ and
$\Lied{\low{\delta_F}}{\low v} = -\low v$. This allows us to infer
$\up v > 0$ and $\low v > 0$ hold at all times.

 We apply the (Sync) rule
%synchronize the dynamics $\up\delta_F$ and $\low\delta_F$
along $\up x = \low x$, yielding the synchronized dynamics
\[
\dot{\up x} = \up v,\,
\dot{\up v} = -\up v^2,\,
\dot{\low x} = \low v \mul (\up v / \low v),\,
\dot{\low v} = -\low v \mul (\up v / \low v)
\;\amp\;
\up v > 0 \AND \low v > 0
\]
Note that the new evolution domain condition $\low v > 0$ allows us to rewrite
$\low v \mul (\up v / \low v)$ to $\up v$.
The synchronizability conditions follow immediately from the fact that
$\low v > 0$ and $\up v > 0$.
For the synchronized formula,
we apply the (DI) rule, so
the desired inequality $\up v \ge \low v$ is reduced to $\up v^2 \le \up v$,
that is, $\up v \le 1$.
To this end,
$\up v > 0$ tells us that the
derivative of $\up v$, that is, $-\up v^2$, is always negative,
therefore $\up v \le 1$.

\subsection{Model Refinement}
\label{sec:switch}

In this example, we consider two abstract models of cars. The first car
is able to provide a high amount of constant acceleration $a$ at low
velocities, but at a certain velocity $v_{cut}$ the engine switches
to a different mode and then provides a lesser, but still constant
acceleration $a_{cut}$. The second car is an abstracted version of the
first, which ignores this mode change and provides the same constant
amount of acceleration $a$ at all velocities. Our aim in this example is 
to establish a safety envelope around the first car's behavior
using the more simply stated second car's dynamics.
Hence we show that the second car's velocity is greater than
the first's at any position $\up x = \low x = l$.
More formally,
the
behavior of the first car is expressed as a hybrid program
$\up \alpha \defeq (\,\up{\delta_1};\ ?\up v=v_{cut};\ \up{\delta_2}\,)$
with two modes: $\up{\delta_1} \defeq (\D{\up x} = \up v, \D{\up v}
= a \amp \up v \le v_{cut})$ and $\up{\delta_2} \defeq (\D{\up x} = \up v,
\D{\up v} = a_{cut})$.
The second car follows the simple dynamics $\low \delta \defeq
\left(\D{\low x} = \low v, \D{\low v} = a\right)$.
Our goal is to prove the sequent
$\Gamma
\vdash \hpleft\, \up \alpha; \low \delta\,\hpright 
(\low x = \up x = l \IMP \low v \le \up v),
$
where the initial conditions are given by
\[
 \Gamma \defeq (\up x = \low x = 0, 0 < \up v = \low v = v_0, 0 < v_{cut}, 0 < a_{cut} \le a)
\]

Technically, the (Sync) rule merges one differential dynamics with
another, but the program the first car executes is a more complicated
composition of dynamics and testing. However, it is possible to
synchronize \emph{piecewise}, first synchronizing $\low \delta$ with $\up{
\delta_1}$ until the first car changes modes, then synchronizing $\low
\delta$ with $\up{\delta_2}$ for the remainder of their runs.
This slightly generalized synchronization procedure means that we can instead
show
\[\Gamma
\vdash \hpleft\, \up{\delta_1} \sync[\up x, \low x] \low \delta; ? \up v = v_{cut};
\up{\delta_2} \sync[\up x, \low x] \low \delta \,\hpright 
(\low x = \up x = l \IMP \low v \le \up v)
\]
There are also now two
sets of synchronizability conditions to satisfy, but both are again
straightforward.
Since $\up{\delta_1}$ and $\low \delta$ are nearly identical (except for
the evolution domain constraint), their synchronization $\up{\delta_1}
\sync[\up x, \low x] \low \delta$ basically identifies the two dynamics.
The synchronization of $\up{\delta_2}$ and $\low \delta$ is exactly the
synchronization performed above in Section~\ref{subsec:const}, and proceeds in the
same way.

\section{Conclusions and Future Work}

In this paper, we present a relational extension of the differential dynamic logic based on
time stretching of dynamics. This reparametrization enables us to
enforce that comparisons between two systems occur when certain conditions
are satisfied, for example when two cars are passing through the same position. 
While such reparametrizations can be
thought of as stretching or compressing time for one of the dynamics,
we also show they can be conducted by a
transformation of the dynamics themselves, based on Lie derivatives.
We call this process \emph{synchronizing} the dynamics
(Definition~\ref{def:sync}), and it leads us to a new $\dL$ proof rule, the
(Sync) rule (Theorem~\ref{thm:sync}). We implemented the new rule 
in the \keymaerax\ tool and use our extension to demonstrate several
nontrivial relational properties of dynamical systems.

In the future, we think it would be interesting to combine our
relational logic with orthogonal relational extensions of $\dL$~\cite{LoosP16} which
focus on \emph{refinement relations} with varying levels of
nondeterminism. We also hinted in our last case study that it is
possible to synchronize wider classes of hybrid programs than just two
differential dynamics. We also think that the level of automated proof
search available in \keymaerax\ may enable the automatic detection of
monotonic properties in \emph{product lines}. This may be useful in
industry both to provide sanity checks on formalized models of
products, as well as enabling strong guarantees to be more easily
obtained for those models.

%\section*{Acknowledgement}
%We are grateful to the KeYmaera X developpers who gave us
%some tips on compilation and modification of the KeYmaera X source code.

% In the end, we found the structure of the original KeYmaera X code to be
% flexible. It easily accommodates our rules---despite the fact that, in our
% rules, Thanks to the way KeYmaera X is implemented, we could implement these
% additional operations simply in Scala.
% comm

 \bibliographystyle{splncs04}
 \bibliography{references}
%
%%%%% To display Open Access text and logo, Please add below text and copy the cc_by_4-0.eps in the manuscript package %%%

\vspace*{\fill}

{\small\medskip\noindent{\bf Open Access} This chapter is licensed under the terms of the Creative Commons\break Attribution 4.0 International License (\url{http://creativecommons.org/licenses/by/4.0/}), which permits use, sharing, adaptation, distribution and reproduction in any medium or format, as long as you give appropriate credit to the original author(s) and the source, provide a link to the Creative Commons license and indicate if changes were made.}

{\small \spaceskip .28em plus .1em minus .1em The images or other third party material in this chapter are included in the chapter's Creative Commons license, unless indicated otherwise in a credit line to the material.~If material is not included in the chapter's Creative Commons license and your intended\break use is not permitted by statutory regulation or exceeds the permitted use, you will need to obtain permission directly from the copyright holder.}

\medskip\noindent\includegraphics{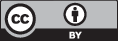}

\end{document}